\documentclass[12pt,a4paper]{article}
\usepackage{amsmath,amssymb,mathrsfs, url,cite,ifpdf,slashed,multirow,tabularx,type1cm}

\ifpdf                                     
  \usepackage{graphicx}                    
\else                                      
  \usepackage[dvipdfmx]{graphicx}          
\fi

\usepackage[colorlinks=true,linkcolor=blue,citecolor=blue,urlcolor=magenta]{hyperref} 

\def\thefootnote{\ifnum\c@footnote>\z@\textasteriskcentered\@arabic\c@footnote\fi}
\makeatletter
\renewcommand{\footnoterule}{%
\kern-3\p@
\hrule width 0.4\columnwidth
\kern 2.6\p@}
\def\thefootnote{\ifnum\c@footnote>\z@\@arabic\c@footnote\fi}
\makeatother

\allowdisplaybreaks

\newcommand{\TeV}{\,{\rm TeV}}
\newcommand{\GeV}{\,{\rm GeV}}

\def\be{\begin{equation}}
\def\ee{\end{equation}}
\def\beq{\begin{eqnarray}}
\def\eeq{\end{eqnarray}}

\def\({\left(}
\def\){\right)}
\def\<{\langle}
\def\>{\rangle}

\makeatletter
\newcommand{\@authornote}[2]{{\def\thefootnote{\fnsymbol{footnote}}\setcounter{footnote}{#1}#2\setcounter{footnote}{0}}}
\newcommand{\authornotemark}[1]{\@authornote#1{\addtocounter{footnote}{-1}\footnotemark}}
\newcommand{\authornotetext}[2]{\@authornote#1{\footnotetext{#2}}}
\makeatother

\begin{document}

\begin{titlepage}

\begin{flushright}
UT--14--26\\
May, 2014
\end{flushright}

\vskip 1.5 cm
\begin{center}

{\Large \bf 
Singlino Resonant Dark Matter and 125~GeV \\[0.5em] Higgs Boson  in High-Scale Supersymmetry}

\vskip .95in

{\large
\textbf{Kazuya Ishikawa}${}^\sharp$ \footnote[0]{${}^\sharp$ {\it E-mail:} \textcolor{magenta}{ishikawa@hep-th.phys.s.u-tokyo.ac.jp}},
\textbf{Teppei Kitahara}${}^\natural$ \footnote[0]{${}^\natural$ {\it E-mail:} \textcolor{magenta}{kitahara@hep-th.phys.s.u-tokyo.ac.jp}},
and
\textbf{Masahiro Takimoto}${}^\flat$ \footnote[0]{${}^\flat$ {\it E-mail:} \textcolor{magenta}{takimoto@hep-th.phys.s.u-tokyo.ac.jp}}
}
\vskip 0.4in

{\large
{\it 
Department of Physics,  Faculty of Science, 
University of Tokyo, \\[0.4em]
Bunkyo-ku, 
Tokyo 113-0033, Japan
}}
\vskip 0.1in

\end{center}
\vskip .65in

\begin{abstract}
We consider a singlino Dark Matter (DM) scenario in a singlet extension model of the Minimal  Supersymmetric  Standard Model, which is so-called the Nearly MSSM (nMSSM).
We find that with high-scale supersymmetry breaking the singlino can obtain a sizable radiative correction to the mass, which opens a window for the  DM scenario with resonant annihilation via the exchange of the Higgs boson.
We show that the current DM relic abundance and the Higgs boson mass can be explained simultaneously. 
This scenario can be fully probed by XENON1T. 
\end{abstract}

\end{titlepage}
\renewcommand{\thefootnote}{\#\arabic{footnote}}
\setcounter{page}{1}
\hrule
\tableofcontents
\vskip .2in
\hrule
\vskip .4in


\section{Introduction}
The supersymmetric (SUSY) models are good candidates of the physics beyond the Standard Model (SM) because they solve the hierarchy problem and ensure the unification of the gauge couplings.
In addition, the lightest SUSY particle (LSP) can be a natural candidate of the Dark Matter (DM) if the R-Parity is conserved.
However the minimal SUSY extension of the SM (MSSM) contains a dimensionful parameter $\mu$ and it causes ``$\mu$-problem''~\cite{Kim:1983dt}. 
Although $\mu$ must be a size of the SUSY breaking scale to realize the electroweak symmetry breaking properly, there is no reason for $\mu$ to be small compared to the Planck scale.
One of the simplest way to solve this problem is introducing a gauge-singlet superfield.
There are several models of singlet extension of the MSSM depending on the imposed symmetry (for a review see~\cite{Ellwanger:2009dp}). 
The Nearly-Minimal (or New Minimal) Supersymmetric Standard Model (nMSSM)~\cite{ Panagiotakopoulos:1999ah, Panagiotakopoulos:2000wp, Dedes:2000jp} based on $\mathbb{Z}^R_5$ or $\mathbb{Z}^R_7$ R-symmetry does not suffer from the domain wall problem, unlike $\mathbb{Z}_3$ symmetric models (NMSSM)~\cite{Abel:1995wk, Panagiotakopoulos:1998yw}.
Therefore, the nMSSM is one of the promising models of the new physics.

On the other hand, recent various cosmological observations have established the $\Lambda$CDM cosmological model and the relic abundance of the cold DM is measured accurately~\cite{Hinshaw:2012aka,Ade:2013zuv}. 
In the nMSSM,  singlino can be a candidate of the DM~\cite{Dedes:2000jp,Menon:2004wv,Barger:2005hb,Balazs:2007pf,Cao:2009ad,Wang:2012ry}.
But they seem to be incompatible with relatively high-scale (TeV scale) supersymmetry breaking, which is inferred from the measured SM Higgs boson mass~\cite{ATLAS:2013mma, CMS:yva} and the null results of the sparticle searches at the Large hadron collider (LHC)~\cite{TheATLAScollaboration:2013fha, CMS:2013cfa}.
This is because the singlino mass and its couplings with SM particles have been thought to be suppressed by the SUSY breaking scale, which leads to the overabundant singlino DM in the universe.
However, if one-loop corrections to the singlino mass are taken into account, the singlino can obtain a sizable mass, which opens a window for a resonant DM scenario via the s-channel annihilation with the exchange of the SM Higgs boson.
Furthermore, in these resonant DM scenarios since the annihilation rate of the singlino is $p$-wave suppressed, one needs a relatively large value of the Higgs-DM coupling.
This fact implies that the singlino DM can be probed more readily than the scalar one~\cite{Kanemura:2010sh}.
 
In this letter, we study the singlino resonant DM scenario within the high-scale nMSSM including one-loop corrections to the neutralino masses.
We will show that if the SUSY breaking scale is around $\sim 10$~TeV and $\tan \beta$ is relatively low, the current DM abundance and the measured SM Higgs boson mass can be achieved simultaneously.
We will also find  that this scenario can be fully probed by the proposed future DM search, XENON1T~\cite{Aprile:2012zx}.

This letter is organized as following.
In section \ref{NMSSM}, we give a short review of the nMSSM.
We present properties of the singlino in section \ref{DMS}.
In section \ref{NumRes}, we investigate the singlino resonant DM scenario with high-scale SUSY breaking, which is compatible with the SM Higgs boson mass $\sim 125$ GeV.
Section \ref{CON} is devoted to the conclusion and discussions.

\section{The Nearly MSSM}
\label{NMSSM}
In this section, we briefly review the nMSSM~\cite{Panagiotakopoulos:1999ah, Panagiotakopoulos:2000wp, Dedes:2000jp}.

In the nMSSM, to solve the $\mu$-problem a gauge-singlet chiral superfield $\hat{S}$ is introduced.
The superpotential and the soft SUSY breaking terms are given as
\begin{align}
	W &= \lambda \hat{S} \hat{H}_{u} \cdot \hat{H}_d + \frac{m_{12}^2}{\lambda} \hat{S} + W_{\textrm{Yukawa}}\,,\\
	\label{hos}
V_{\rm soft}&=m_S^2|S|^2+
\left( \lambda A_\lambda H_u\cdot H_d S+t_S S+\textrm{h.c.} \right) +V_\textrm{soft}^\textrm{MSSM}\,,
\end{align}
where $\hat{H}_u$ ($\hat{H}_d$) is up(down)-type 
Higgs doublet superfield.
Although the terms $m_{12}^2$ and $t_S$ are forbidden by a discrete $\mathbb{Z}^R_5$ ($\mathbb{Z}^R_7$) R-symmetry when supersymmetry is conserved, they are generated by supergravity effects as
\beq
m_{12}^2 & = & \lambda \xi_F M_S^2\,, \\
t_S & = &\xi_S M_S^3\,, 
\eeq
where $M_S$ denotes the SUSY breaking scale (see Refs.~\cite{Panagiotakopoulos:1999ah, Panagiotakopoulos:2000wp, Dedes:2000jp}). 
Here $\xi_F$ and $\xi_S$ are $\mathcal{O}(1)$ constants and then $m_{12}^2$ and $t_S$ become $O(M_S^2)$ and $O(M_S^3)$ respectively \footnote{
Although the trilinear $\kappa S^3$ term is also generated, it is highly suppressed by
 Planck scale.}.
With these values, $S$ has a vacuum expectation value $\langle S \rangle \sim -t_S/m_S^2\sim O(M_S)$.
This vacuum expectation value generates an effective $\mu$-parameter $\mu_{\textrm{eff}}\equiv\lambda \langle S \rangle \sim O(M_S)$ and $\mu$-problem is solved.

At the tree level, the neutralino mass matrix in the basis 
$(\tilde{B}, \tilde{W}^0, \tilde{H}_d^0, \tilde{H}_u^0, \tilde{S})$ is
\begin{align}
\label{NMASS}
	\mathcal{M}_\text{tree}=
       \begin{pmatrix}	
	M_1 & 0 & -\frac{g_1 v_d}{\sqrt{2}}& \frac{g_1 v_u}{\sqrt{2}} &0\\
	 ~& M_2 &  \frac{g_2 v_d}{\sqrt{2}}& -\frac{g_2 v_u}{\sqrt{2}} &0\\
	 ~&~&0&-\mu_{\rm eff}&-\lambda v_u\\
	 ~&~&~&0&-\lambda v_d\\
	 ~&~&~&~&0
	 \end{pmatrix}\,,
\end{align}
where $\tilde{B}$ is the bino, $\tilde{W}^0$ is the neutral wino, $\tilde{H}_d^0$ and $\tilde{H}_u^0$ are the neutral Higgsinos and $\tilde{S}$ is the fermionic component of $\hat{S}$.
$v_u$ ($v_d$) is the vacuum expectation value of $H^0_u$ ($H_d^0$) with $v^2\equiv v_u^2+v_d^2\simeq (174\text{ GeV})^2$. 
$M_1$ and $M_2$ are the gaugino masses, where the gauge couplings for U(1)$_Y$ and SU(2) are denoted as $g_1$ and $g_2$ respectively.
We denote $\tilde{s}$ as the mass-eigenstate neutralino whose component is mainly $\tilde{S}$. 
We call $\tilde{s}$ as a singlino in this letter.
When the SUSY breaking scale is relatively high as suggested by the LHC experiments~\cite{ATLAS:2013mma, CMS:yva, TheATLAScollaboration:2013fha, CMS:2013cfa}, the singlino becomes the LSP and it can be a good candidate of the DM.

In the nMSSM, since the SM Higgs boson has an extra contribution to the quartic coupling $\lambda_\textrm{quartic}$, there is a sizable tree-level contribution to the Higgs boson mass.
When integrating out heavy SUSY particles and matching with the SM, the SM Higgs quartic coupling is shifted by~\cite{Giudice:2011cg}
\beq
\label{masscor}
\delta \lambda_{\textrm{quartic}} = \frac{\lambda^2 }{2} \frac{m_S^2 - A_{\lambda}^2}{m_S^2} \sin^2 2 \beta\,,
\eeq
compared to the MSSM.
Large $\lambda $ and small $\tan \beta$ can give a sizable contribution to the Higgs boson mass.
However, note that this extra contribution becomes small if $m_S \sim A_{\lambda}$.

\section{DM abundance and Radiative Singlino mass in the nMSSM}
\label{DMS}
In this section, we calculate the DM abundance and briefly estimate the singlino mass in the nMSSM.

Let us consider the case where only the singlino $\tilde{s}$ is light and other SUSY particles are relatively heavy. 
In this case, the low energy effective Lagrangian can be written as
\begin{align}
\label{low}
	-\mathcal{L}_{\rm eff}\supset \frac{m_{\tilde{s}}}{2}\bar{\tilde{s}}\tilde{s}
	+\frac{\lambda_{\rm eff}}{2}h\bar{\tilde{s}}\tilde{s}\,, 
\end{align}
where $h$ corresponds to the SM Higgs boson.
Before going to the numerical calculation in the nMSSM, we estimate the thermal relic abundance of singlino with this effective model regarding $\lambda_{\rm eff}$ and $m_{\tilde{s}}$ as free parameters by solving Bolzmann equation~\cite{Gondolo:1990dk}.
\begin{figure}[t]
\begin{center}
\includegraphics[width =12cm]{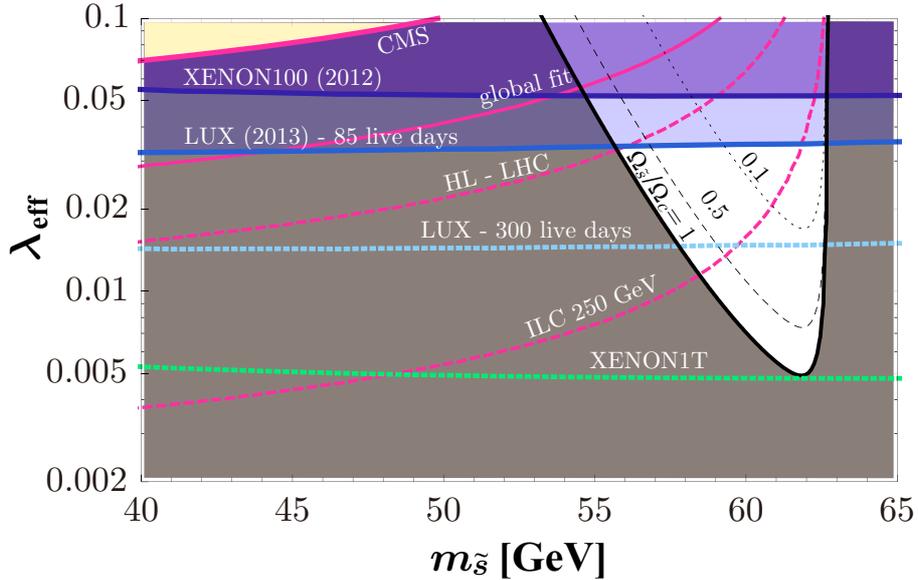}
\caption{The singlino thermal relic abundance and experimental constraints/future prospects.
The black lines denote the ratio of the thermal relic abundance $\Omega_{\tilde{s}} h^2$ to the current DM density $\Omega_c h^2 = 0.1199$~\cite{Ade:2013zuv}.
The singlino relic density overclose the universe at the dark-shaded region.
The regions above the red solid lines are excluded by the Higgs invisible decay ($h \to \tilde{s} \tilde{s}$) searches of CMS (Br${}_h^{\textrm{inv.}}$ $\leq$ 58~\%) \cite{Chatrchyan:2014tja} for upper line (yellow-shaded region) and by the global fit of the Higgs couplings ($19~\%$)~\cite{Belanger:2013xza} for lower line.
The dashed red lines correspond to the future sensitivity of high luminosity LHC (6.2~\%)~\cite{Dawson:2013bba} and ILC with $\mathcal{L} = 1150 \textrm{fb}{}^{-1}$ at $\sqrt{s} = 250 \GeV$ (0.4~\%) \cite{Asner:2013psa}.
The blue-shaded regions are excluded by XENON100~\cite{Aprile:2012nq} and LUX~\cite{Akerib:2013tjd}.
The regions above the blue and the green dashed lines can be probed by the future direct DM searches of LUX~\cite{Akerib:2012ys} and XENON1T~\cite{Aprile:2012zx}.
}\label{abundance}
\end{center}
\end{figure}
In Fig.~\ref{abundance}, the black lines show the ratio of the thermal relic abundance $\Omega_{\tilde{s}} h^2$ to the current DM density $\Omega_c h^2 = 0.1199$~\cite{Ade:2013zuv} where we take the Higgs boson mass as $m_h=125.5$ GeV.
The regions above the red solid lines are excluded by the Higgs invisible decay ($h \to \tilde{s} \tilde{s}$) searches of CMS (upper line)~\cite{Chatrchyan:2014tja} and by the global fit of the Higgs couplings (lower line)~\cite{Belanger:2013xza}.
The regions above the red dashed lines can be probed by the future Higgs invisible decay searches of high luminosity LHC (upper line)~\cite{Dawson:2013bba} and ILC (lower line)~\cite{Asner:2013psa}.
The direct DM searches set limits on the spin-independent cross section of DM-nucleon elastic scattering.
The blue-shaded regions are excluded by the direct DM searches of XENON100~\cite{Aprile:2012nq} and LUX~\cite{Akerib:2013tjd}.
The region above the blue (green) dashed line can be probed by the future direct DM search of LUX~\cite{Akerib:2012ys} (XENON1T~\cite{Aprile:2012zx}).
For applying these constraints and future prospects, we assume $\Omega_{\tilde{s}} h^2 = \Omega_c h^2$.
The gray-shaded region is excluded by the overclosure of the universe.
One can see that the region where $\tilde{s}$ is consistent with the current DM relic abundance lies around $\lambda_{\rm eff}\sim \mathcal{O}(0.01)$ and ${m_{\tilde{s}}}\sim 60$ GeV. 
In this region, resonant pair-annihilation of $\tilde{s}$ occurs via the Higgs boson with $m_{\tilde{s}} \sim m_h/2$.
This allowed region can be covered by the future Higgs invisible decay searches and direct DM searches, especially by XENON1T.

\begin{figure}[t]
\begin{center}
\includegraphics[width =6cm]{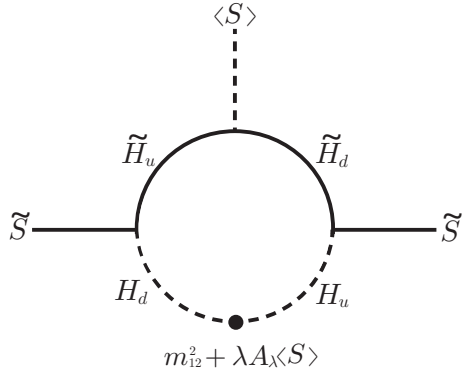}
\caption{Typical one-loop diagram which contributes to the singlino mass. 
}\label{diagram}
\end{center}
\end{figure}

Now, we estimate $m_{\tilde{s}}$ and $\lambda_{\rm eff}$ in the nMSSM.
From the tree-level calculations, these values are evaluated as
\begin{align}
	m_{\tilde{s}}^\textrm{tree}\sim\lambda^2 \frac{v^2}{M_S} \sin{2\beta}\,,\\ \label{treecoup}
	\lambda_\textrm{eff}^\textrm{tree}\sim \lambda^2 \frac{v}{M_S}\sin{2\beta}\,, 
\end{align}
where $\tan \beta\equiv {v_u}/{v_d}$ and we denotes the typical SUSY breaking scale by $M_S$.
Obviously $\lambda_{\rm eff}\sim \mathcal{O}(0.01)$ and $m_{\tilde{s}}\sim 60$ GeV can not be satisfied at the same time.
However, one-loop corrections to the neutralino mass~\cite{Staub:2010ty} can raise the singlino mass with relatively large $M_S$.
The typical diagram which contributes to the singlino mass is given in Fig.~\ref{diagram}.
The one-loop singlino mass can be estimated as
\begin{align} \label{oneloopmass}
	m_{\tilde{s}}^\textrm{1-loop}&\sim
	\frac{\lambda^2}{(4\pi)^2}\mu_{\textrm{eff}}\sin2\beta \cdot F\left( \frac{2 (m_{12}^2 + A_{\lambda} \mu_{\textrm{eff}})}{\mu_{\textrm{eff}}^2 \sin 2 \beta}\right) \nonumber \\
	&\sim
	\frac{\lambda^2}{(4\pi)^2}M_S\sin 2\beta \,,
\end{align}
where the loop function $F\left( x\right)$ is defined as $F\left(x\right) \equiv (x \log x)/(x-1)$ and satisfies $F\left(1\right) = 1$.
We calculate the singlino mass including the full one-loop corrections~\cite{Staub:2010ty} \footnote{In the limit of $\kappa=0$, one-loop corrections in the NMSSM reduce to the one in the nMSSM. 
We found that Ref.~\cite{Staub:2010ty} includes some typos in the equations of the one-loop corrections~\cite{private}.}.
\begin{figure}[ht]
\begin{center}
\includegraphics[width =12cm]{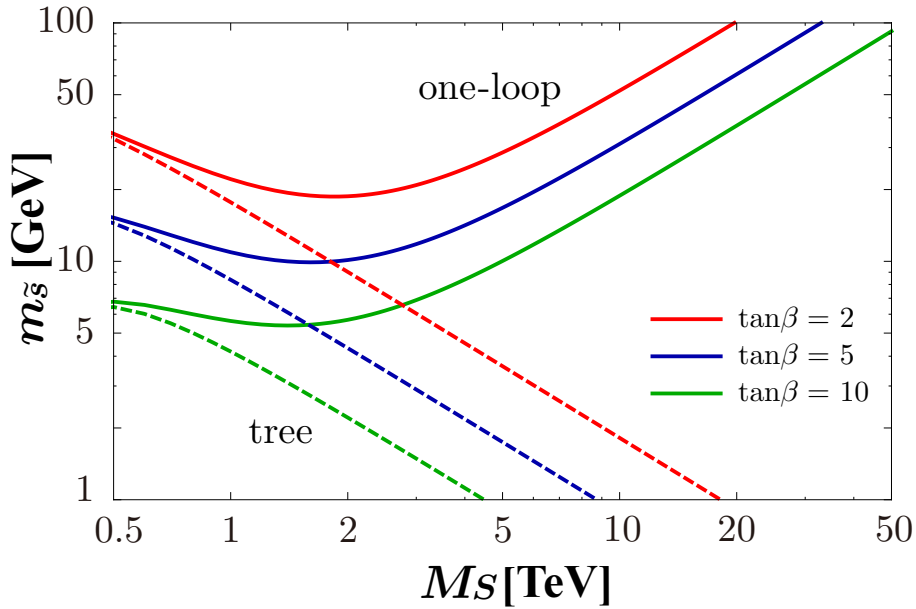}
\caption{The one-loop-level singlino mass and the tree-level one as a function of $M_S$. 
}\label{singlinomass}
\end{center}
\end{figure}
Fig.~\ref{singlinomass} shows the dependence of $M_S$ to the singlino mass in the tree level and the one-loop level. 
In this figure, we take $\lambda = 0.75$ and all dimensionful parameters equal to $M_S$.
One can see that the singlino obtains sizable one-loop corrections to the mass in high-scale SUSY scenario.
Since this feature is due to the suppression of the singlino mass at the tree level, the two-loop level corrections to the singlino mass is estimated to be smaller than the one-loop one.
Note that with $M_S\sim O(10)$ TeV, $\tan\beta  \sim O(1)$ and $\lambda \sim O(1)$, one can simply obtain $\lambda_{\rm eff}\sim \mathcal{O}(0.01)$ and $m_{\tilde{s}}\sim 60$ GeV \footnote{The one-loop $\lambda_{\textrm{eff}}$ can be roughly estimated as $\lambda_\textrm{eff}^\textrm{1-loop} \sim \frac{\lambda^4}{(4 \pi)^2} \frac{v}{M_S} \sin 2 \beta$, which is negligible in comparison with $\lambda_\textrm{eff}^\textrm{tree}$.}.
Moreover, the Higgs boson mass becomes around $125$ GeV in such parameter sets with the help of the additional quartic coupling $\lambda$.
We show these validity by using the numerical calculations in the next section.

\section{Numerical Results}
\label{NumRes}
In this section, we numerically investigate the singlino resonant DM scenario and the Higgs boson mass in the nMSSM.
In this letter, we calculate the Higgs boson mass using the two-loop renormalization group equation including the matching condition (\ref{masscor})~\cite{Giudice:2011cg}.

\begin{figure}[t]
\begin{center}
\includegraphics[width =13cm]{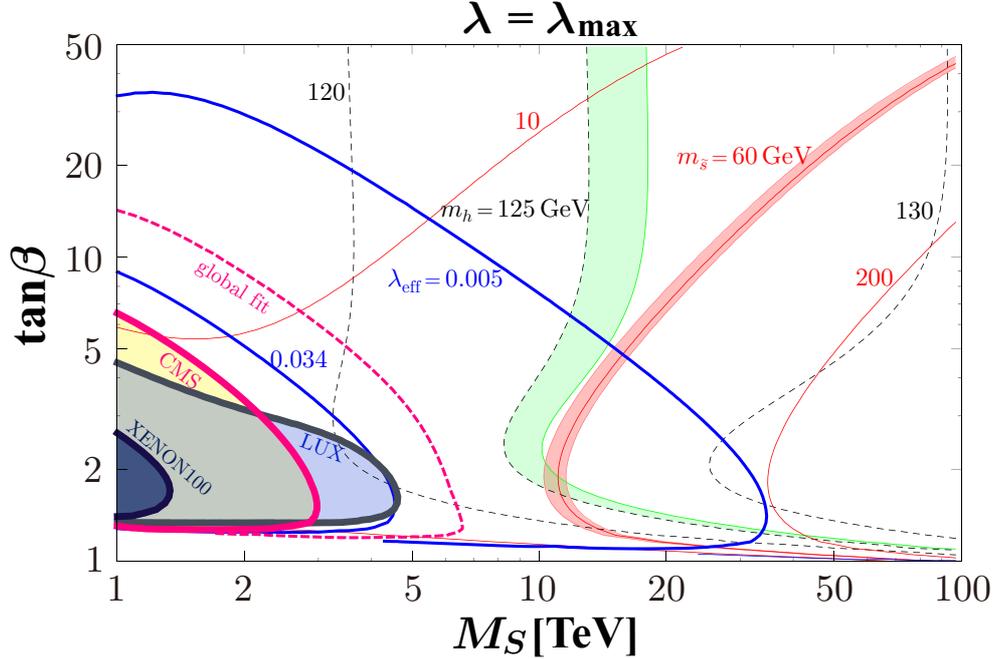}
\caption{Contours of $m_{\tilde{s}}$ (red lines), $\lambda_{\textrm{eff}}$ (blue lines) and $m_h$ (black dashed lines) in $M_S$-$\tan\beta$ plane assuming $\lambda = \lambda_{\textrm{max}}$ at each point.
On the red-shaded region ( $55.5 \GeV < m_{\tilde{s}} < 62.7 \GeV$ ), the resonant annihilation via the Higgs boson can occur.
The green-shaded region satisfies $125 \GeV < m_h < 126 \GeV$.
The blue (dark blue)-shaded region is excluded by the current limits from LUX~\cite{Akerib:2013tjd} (XENON~\cite{Aprile:2012nq}).
The yellow-shaded region is excluded by the Higgs invisible decay search at the CMS~\cite{Chatrchyan:2014tja} and the magenta dashed line is the current bound by the global fit of the Higgs coulings~\cite{Belanger:2013xza}.
}\label{lambdamax}
\end{center}
\end{figure}

In Fig.~\ref{lambdamax}, we show the singlino mass $m_{\tilde{s}}$ (red lines), the effective Higgs-DM coupling $\lambda_{\textrm{eff}}$ (blue lines) and the Higgs boson mass $m_h$ (black dashed lines) in $M_S$-$\tan\beta$ plane.
For simplicity, all parameters are chosen to be real.
The trilinear coupling $\lambda$ is taken to be $\lambda_{\textrm{max}}$ which is a maximal value avoiding Landau singularities up to  the GUT scale, $2\times10^{16}\GeV$.
All SUSY breaking parameters except $A_\lambda$ are set to $M_S$ ($\lambda  \xi_F = \xi_S = 1$).
In order to obtain a sizable contribution to the Higgs boson mass, we choose $A^2_\lambda = \frac{2}{5} M_S^2$.
As one can see from Fig.~\ref{abundance}, the viable regions for the singlino DM are $55.5 \GeV < m_{\tilde{s}} < 62.7 \GeV$ and  $0.005 < \lambda_{\textrm{eff}} < 0.034 $.
In Fig.~\ref{lambdamax}, these regions correspond to the red-shaded band and the region between the two blue lines respectively. 
The green band represents $125 \GeV < m_h < 126 \GeV$.
One can see that the singlino resonant DM scenario is successful with $\tan\beta \sim \mathcal{O}(1)$ and $M_S \sim \mathcal{O}(10) \TeV$. 

If we choose the lower value of $A^2_\lambda$, the green line moves to left because the Higgs boson mass obtains more contribution from the quartic coupring (see Eq.~(\ref{masscor})). 
On the other hand, with smaller value of $m_{12}^2+\lambda A_\lambda \langle S\rangle$ the singlino mass becomes lighter and the red-shaded region moves to right.
The blue lines are not sensitive to the choice of $m_{12}^2$ and $A_\lambda$, because $\lambda_{\textrm{eff}}$ is determined by the SUSY breaking scale and $\tan\beta$.
The important point is that in any case with $M_S\sim \mathcal{O}(10)$ TeV and low $\tan \beta$ the current DM abundance and the measured Higgs boson mass can be realized simultaneously.
This opens a window for the singlino DM in high-scale supersymmetry.

\begin{figure}[t]
\begin{center}
\includegraphics[width =13cm]{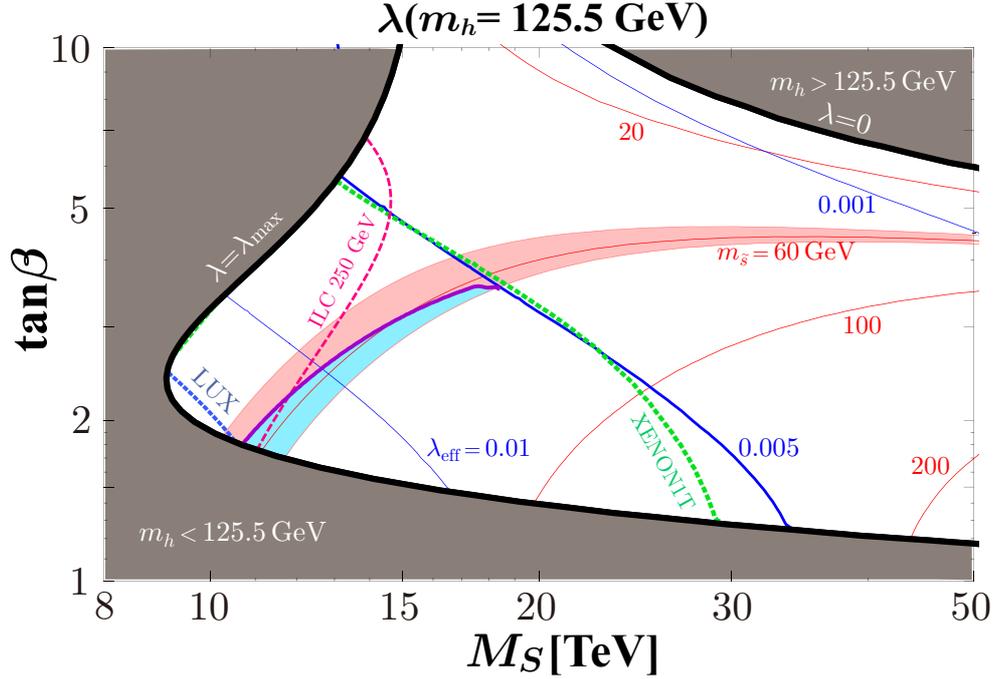}
\caption{Contours of $m_{\tilde{s}}$ (red lines), $\lambda_{\textrm{eff}}$ (blue lines) in $M_S$-$\tan\beta$ plane under $m_h = 125.5 \GeV$ by changing $\lambda$, $0 \leq \lambda \leq \lambda_\textrm{max}$.
On the purple line, the singlino relic abundance $\Omega_{\tilde{s}} h^2$ is consistent with the current value, $\Omega_c h^2 = 0.1199$~\cite{Ade:2013zuv}. 
In the light blue region, $\Omega_{\tilde{s}}h^2 \leq \Omega_c h^2$. 
The left side of the blue (green) dashed line can be probed by the future DM search LUX~\cite{Akerib:2012ys} (XENON1T~\cite{Aprile:2012zx}).
ILC~\cite{Asner:2013psa} can cover the left side of the magenta dashed line.
Other lines are the same in Fig.~\ref{lambdamax}.
}\label{higgs1255}
\end{center}
\end{figure}

Finally, we show these regions in detail (see Fig.~\ref{higgs1255}).
In this figure, the Higgs boson mass is fixed to be $125.5$ GeV by changing $\lambda$, $0 \leq \lambda \leq \lambda_\textrm{max}$.
The input parameters are the same as Fig.~\ref{lambdamax} except $\lambda$.
In the dark-shaded regions, one can not explain $m_h = 125.5\GeV$. 
The singlino relic abundance $\Omega_{\tilde{s}} h^2$ is consistent with the current value on the purple line, $\Omega_c h^2 = 0.1199$~\cite{Ade:2013zuv}. 
In the light blue region, $\Omega_{\tilde{s}}h^2 \leq \Omega_c h^2$. 
The left side of  the dashed lines can be covered by LUX (blue) \cite{Akerib:2012ys}, XENON1T (green) \cite{Aprile:2012zx} and  ILC (magenta) \cite{Asner:2013psa}.
From this result, the future experiments can probe a sign of the singlino  DM.

\section{Conclusion and Discussions}
\label{CON}
In this letter, we have studied the singlino resonant DM scenario within the high-scale nMSSM.
Including one-loop corrections to the neutralino masses, the singlino can explain the current DM relic abundance through the resonant annihilation via the Higgs boson.
We have shown that with high-scale SUSY breaking $\sim 10$ TeV and low $\tan\beta$, the  DM relic abundance and the SM Higgs boson mass can be explained simultaneously in this scenario.
Even for the high-scale SUSY, we have also shown that the parameter region where the singlino DM is consistent with the current DM relic abundance can be fully probed by the future experiments (see Fig.~\ref{abundance},~\ref{higgs1255}).
Therefore, the singlino DM signal can be {\it ``a first sign"} of  the high-scale supersymmetry.

In this papar, we have concentrated on the scenario of singlino DM with resonant annihilation via the Higgs boson. 
Now, let us consider other scenarios of the high-scale nMSSM.
After integrating out heavy SUSY particles, in the effective theory there are SM particles and only one additional particle, singlino.
The singlino has interactions with the SM Higgs boson and with the Z boson.
While the effective coupling with the Higgs boson is suppressed by $v/M_S$, the coupling with the Z boson is more suppressed by $\sim (v/M_S)^2$, which prevent the resonant scenario with the Z boson.
In order for the singlino not to be overabundant, the resonant scenario with the SM Higgs boson is the last resort for the high-scale nMSSM.

The NMSSM is another model of the singlet extension of MSSM~\cite{Ellwanger:2009dp}.
The superpotential is given as
\beq
W_{\textrm{NMSSM}} = \lambda \hat{S} \hat{H_u} \cdot \hat{H}_d + \frac{\kappa}{3} \hat{S}^3 + W_{\textrm{Yukawa}}.
\eeq
In the NMSSM, the singlino can obtain a radiative correction to the mass in addition to the tree-level mass $m_{\tilde{s}}^\textrm{tree} \sim 2 \kappa \langle S \rangle$.
The singlino resonant DM scenario may be successful with small $\tan\beta$  and small $\kappa$ in high-scale SUSY scenario.
In the small $\kappa$ limit, a singlet-like CP-odd scalar boson $a$ becomes a pseudo Nambu-Goldstone boson because of the existence of the global U(1) Peccei-Quinn symmetry.
Therefore, one may be able to make a distinction between the singlino resonant scenario in the nMSSM and NMSSM by the search for $h \to aa $~\cite{Cao:2013gba}.
 
Since there are some new sources of CP violating phases in the nMSSM, the electric dipole moment (EDM) are generally generated through relative phase between $\mu_{\textrm{eff}}$ and $M_{\textrm{gaugino}}$ at the one-loop level.
The electron EDM is roughly evaluated as 
\beq
\left|\frac{d_e}{e} \right| &\sim& \frac{5 g_2^2 + g_1^2 }{384 \pi^2} \frac{m_e}{M_S^2} \sin\phi \tan \beta  \hspace{0.6em}[\GeV^{-1}] \nonumber \\
&\sim& 6 \times 10^{-29} \left( \frac{10\TeV}{M_S} \right)^2 \sin\phi \tan\beta \hspace{0.6em}[\textrm{cm}],
\eeq
where $\phi = \textrm{arg}\left(\mu_{\textrm{eff}} M_{\textrm{gaugino}}\right)$.
One can obtain  $|d_e| \sim \mathcal{O}(10^{-29})$ $e~\textrm{cm}$ with $\tan\beta \sim \mathcal{O}(1)$, $M_S \sim \mathcal{O}(10) \TeV$ and $\sin \phi \sim \mathcal{O}(1)$.
Interestingly, the electron EDM of this size does not conflict with the current bound~\cite{Baron:2013eja} and can be probed by some future experiments~\cite{Sakemi:2011zz, Kawall:2011zz, Kara:2012ay}.

\section*{Acknowledgments}
The authors would like to thank Florian~Staub for providing the fixed  one-loop correction sets of the NMSSM.
They are also grateful to Motoi Endo and Kazunori Nakayama for useful comments and discussions. 
The work of M.T. is supported in part by JSPS Research Fellowships for Young Scientists. The work of M.T. is also supported by the program for Leading Graduate Schools, MEXT, Japan.

\bibliography{Singlino.bib}

\providecommand{\href}[2]{#2}\begingroup\raggedright\begin{thebibliography}{10}

\bibitem{Kim:1983dt}
J.~E. Kim and H.~P. Nilles, ``{The mu Problem and the Strong CP Problem},''
\href{http://dx.doi.org/10.1016/0370-2693(84)91890-2}{{\em Phys.Lett.}
  {\bfseries B138} (1984) 150}.

\bibitem{Ellwanger:2009dp}
U.~Ellwanger, C.~Hugonie, and A.~M. Teixeira, ``{The Next-to-Minimal
  Supersymmetric Standard Model},''
  \href{http://dx.doi.org/10.1016/j.physrep.2010.07.001}{{\em Phys.Rept.}
  {\bfseries 496} (2010) 1--77},
\href{http://arxiv.org/abs/0910.1785}{{\ttfamily arXiv:0910.1785 [hep-ph]}}.

\bibitem{Panagiotakopoulos:1999ah}
C.~Panagiotakopoulos and K.~Tamvakis, ``{New minimal extension of MSSM},''
  \href{http://dx.doi.org/10.1016/S0370-2693(99)01247-2}{{\em Phys.Lett.}
  {\bfseries B469} (1999) 145--148},
\href{http://arxiv.org/abs/hep-ph/9908351}{{\ttfamily arXiv:hep-ph/9908351
  [hep-ph]}}.

\bibitem{Panagiotakopoulos:2000wp}
C.~Panagiotakopoulos and A.~Pilaftsis, ``{Higgs scalars in the minimal
  nonminimal supersymmetric standard model},''
  \href{http://dx.doi.org/10.1103/PhysRevD.63.055003}{{\em Phys.Rev.}
  {\bfseries D63} (2001) 055003},
\href{http://arxiv.org/abs/hep-ph/0008268}{{\ttfamily arXiv:hep-ph/0008268
  [hep-ph]}}.

\bibitem{Dedes:2000jp}
A.~Dedes, C.~Hugonie, S.~Moretti, and K.~Tamvakis, ``{Phenomenology of a new
  minimal supersymmetric extension of the standard model},''
  \href{http://dx.doi.org/10.1103/PhysRevD.63.055009}{{\em Phys.Rev.}
  {\bfseries D63} (2001) 055009},
\href{http://arxiv.org/abs/hep-ph/0009125}{{\ttfamily arXiv:hep-ph/0009125
  [hep-ph]}}.

\bibitem{Abel:1995wk}
S.~Abel, S.~Sarkar, and P.~White, ``{On the cosmological domain wall problem
  for the minimally extended supersymmetric standard model},''
  \href{http://dx.doi.org/10.1016/0550-3213(95)00483-9}{{\em Nucl.Phys.}
  {\bfseries B454} (1995) 663--684},
\href{http://arxiv.org/abs/hep-ph/9506359}{{\ttfamily arXiv:hep-ph/9506359
  [hep-ph]}}.

\bibitem{Panagiotakopoulos:1998yw}
C.~Panagiotakopoulos and K.~Tamvakis, ``{Stabilized NMSSM without domain
  walls},'' \href{http://dx.doi.org/10.1016/S0370-2693(98)01493-2}{{\em
  Phys.Lett.} {\bfseries B446} (1999) 224--227},
\href{http://arxiv.org/abs/hep-ph/9809475}{{\ttfamily arXiv:hep-ph/9809475
  [hep-ph]}}.

\bibitem{Hinshaw:2012aka}
{\bfseries WMAP} Collaboration, G.~Hinshaw {\em et~al.}, ``{Nine-Year Wilkinson
  Microwave Anisotropy Probe (WMAP) Observations: Cosmological Parameter
  Results},'' \href{http://dx.doi.org/10.1088/0067-0049/208/2/19}{{\em
  Astrophys.J.Suppl.} {\bfseries 208} (2013) 19},
\href{http://arxiv.org/abs/1212.5226}{{\ttfamily arXiv:1212.5226
  [astro-ph.CO]}}.

\bibitem{Ade:2013zuv}
{\bfseries Planck} Collaboration, P.~Ade {\em et~al.}, ``{Planck 2013 results.
  XVI. Cosmological parameters},''
\href{http://arxiv.org/abs/1303.5076}{{\ttfamily arXiv:1303.5076
  [astro-ph.CO]}}.

\bibitem{Menon:2004wv}
A.~Menon, D.~Morrissey, and C.~Wagner, ``{Electroweak baryogenesis and dark
  matter in the nMSSM},''
  \href{http://dx.doi.org/10.1103/PhysRevD.70.035005}{{\em Phys.Rev.}
  {\bfseries D70} (2004) 035005},
\href{http://arxiv.org/abs/hep-ph/0404184}{{\ttfamily arXiv:hep-ph/0404184
  [hep-ph]}}.

\bibitem{Barger:2005hb}
V.~Barger, P.~Langacker, and H.-S. Lee, ``{Lightest neutralino in extensions of
  the MSSM},'' \href{http://dx.doi.org/10.1016/j.physletb.2005.09.023}{{\em
  Phys.Lett.} {\bfseries B630} (2005) 85--99},
\href{http://arxiv.org/abs/hep-ph/0508027}{{\ttfamily arXiv:hep-ph/0508027
  [hep-ph]}}.

\bibitem{Balazs:2007pf}
C.~Balazs, M.~S. Carena, A.~Freitas, and C.~Wagner, ``{Phenomenology of the
  nMSSM from colliders to cosmology},''
  \href{http://dx.doi.org/10.1088/1126-6708/2007/06/066}{{\em JHEP} {\bfseries
  0706} (2007) 066},
\href{http://arxiv.org/abs/0705.0431}{{\ttfamily arXiv:0705.0431 [hep-ph]}}.

\bibitem{Cao:2009ad}
J.~Cao, H.~E. Logan, and J.~M. Yang, ``{Experimental constraints on nMSSM and
  implications on its phenomenology},''
  \href{http://dx.doi.org/10.1103/PhysRevD.79.091701}{{\em Phys.Rev.}
  {\bfseries D79} (2009) 091701},
\href{http://arxiv.org/abs/0901.1437}{{\ttfamily arXiv:0901.1437 [hep-ph]}}.

\bibitem{Wang:2012ry}
W.~Wang, ``{A comparative study of dark matter in the MSSM and its singlet
  extensions: a mini review},''
  \href{http://dx.doi.org/10.1155/2012/216941}{{\em Adv.High Energy Phys.}
  {\bfseries 2012} (2012) 216941},
\href{http://arxiv.org/abs/1205.5081}{{\ttfamily arXiv:1205.5081 [hep-ph]}}.

\bibitem{ATLAS:2013mma}
{\bfseries ATLAS} Collaboration,
``{Combined measurements of the mass and signal strength of the Higgs-like
  boson with the ATLAS detector using up to 25 fb$^{-1}$ of proton-proton
  collision data},''.

\bibitem{CMS:yva}
{\bfseries CMS} Collaboration,
``{Combination of standard model Higgs boson searches and measurements of the
  properties of the new boson with a mass near 125 GeV},''.

\bibitem{TheATLAScollaboration:2013fha}
{\bfseries ATLAS} Collaboration,
``{Search for squarks and gluinos with the ATLAS detector in final states with
  jets and missing transverse momentum and 20.3 fb$^{-1}$ of $\sqrt{s}=8$ TeV
  proton-proton collision data},''.

\bibitem{CMS:2013cfa}
{\bfseries CMS} Collaboration,
``{Search for supersymmetry using razor variables in events with b-jets in pp
  collisions at 8 TeV},''.

\bibitem{Kanemura:2010sh}
S.~Kanemura, S.~Matsumoto, T.~Nabeshima, and N.~Okada, ``{Can WIMP Dark Matter
  overcome the Nightmare Scenario?},''
  \href{http://dx.doi.org/10.1103/PhysRevD.82.055026}{{\em Phys.Rev.}
  {\bfseries D82} (2010) 055026},
\href{http://arxiv.org/abs/1005.5651}{{\ttfamily arXiv:1005.5651 [hep-ph]}}.

\bibitem{Aprile:2012zx}
{\bfseries XENON1T} Collaboration, E.~Aprile, ``{The XENON1T Dark Matter Search
  Experiment},''
\href{http://arxiv.org/abs/1206.6288}{{\ttfamily arXiv:1206.6288
  [astro-ph.IM]}}.

\bibitem{Giudice:2011cg}
G.~F. Giudice and A.~Strumia, ``{Probing High-Scale and Split Supersymmetry
  with Higgs Mass Measurements},''
  \href{http://dx.doi.org/10.1016/j.nuclphysb.2012.01.001}{{\em Nucl.Phys.}
  {\bfseries B858} (2012) 63--83},
\href{http://arxiv.org/abs/1108.6077}{{\ttfamily arXiv:1108.6077 [hep-ph]}}.

\bibitem{Gondolo:1990dk}
P.~Gondolo and G.~Gelmini, ``{Cosmic abundances of stable particles: Improved
  analysis},''
\href{http://dx.doi.org/10.1016/0550-3213(91)90438-4}{{\em Nucl.Phys.}
  {\bfseries B360} (1991) 145--179}.

\bibitem{Chatrchyan:2014tja}
{\bfseries CMS} Collaboration, S.~Chatrchyan {\em et~al.}, ``{Search for
  invisible decays of Higgs bosons in the vector boson fusion and associated ZH
  production modes},''
\href{http://arxiv.org/abs/1404.1344}{{\ttfamily arXiv:1404.1344 [hep-ex]}}.

\bibitem{Belanger:2013xza}
G.~Belanger, B.~Dumont, U.~Ellwanger, J.~Gunion, and S.~Kraml, ``{Global fit to
  Higgs signal strengths and couplings and implications for extended Higgs
  sectors},'' \href{http://dx.doi.org/10.1103/PhysRevD.88.075008}{{\em
  Phys.Rev.} {\bfseries D88} (2013) 075008},
\href{http://arxiv.org/abs/1306.2941}{{\ttfamily arXiv:1306.2941 [hep-ph]}}.

\bibitem{Dawson:2013bba}
S.~Dawson, A.~Gritsan, H.~Logan, J.~Qian, C.~Tully, {\em et~al.}, ``{Working
  Group Report: Higgs Boson},''
\href{http://arxiv.org/abs/1310.8361}{{\ttfamily arXiv:1310.8361 [hep-ex]}}.

\bibitem{Asner:2013psa}
D.~Asner, T.~Barklow, C.~Calancha, K.~Fujii, N.~Graf, {\em et~al.}, ``{ILC
  Higgs White Paper},''
\href{http://arxiv.org/abs/1310.0763}{{\ttfamily arXiv:1310.0763 [hep-ph]}}.

\bibitem{Aprile:2012nq}
{\bfseries XENON100} Collaboration, E.~Aprile {\em et~al.}, ``{Dark Matter
  Results from 225 Live Days of XENON100 Data},''
  \href{http://dx.doi.org/10.1103/PhysRevLett.109.181301}{{\em Phys.Rev.Lett.}
  {\bfseries 109} (2012) 181301},
\href{http://arxiv.org/abs/1207.5988}{{\ttfamily arXiv:1207.5988
  [astro-ph.CO]}}.

\bibitem{Akerib:2013tjd}
{\bfseries LUX} Collaboration, D.~Akerib {\em et~al.}, ``{First results from
  the LUX dark matter experiment at the Sanford Underground Research
  Facility},'' \href{http://dx.doi.org/10.1103/PhysRevLett.112.091303}{{\em
  Phys.Rev.Lett.} {\bfseries 112} (2014) 091303},
\href{http://arxiv.org/abs/1310.8214}{{\ttfamily arXiv:1310.8214
  [astro-ph.CO]}}.

\bibitem{Akerib:2012ys}
{\bfseries LUX} Collaboration, D.~Akerib {\em et~al.}, ``{The Large Underground
  Xenon (LUX) Experiment},''
  \href{http://dx.doi.org/10.1016/j.nima.2012.11.135}{{\em Nucl.Instrum.Meth.}
  {\bfseries A704} (2013) 111--126},
\href{http://arxiv.org/abs/1211.3788}{{\ttfamily arXiv:1211.3788
  [physics.ins-det]}}.

\bibitem{Staub:2010ty}
F.~Staub, W.~Porod, and B.~Herrmann, ``{The Electroweak sector of the NMSSM at
  the one-loop level},'' \href{http://dx.doi.org/10.1007/JHEP10(2010)040}{{\em
  JHEP} {\bfseries 1010} (2010) 040},
\href{http://arxiv.org/abs/1007.4049}{{\ttfamily arXiv:1007.4049 [hep-ph]}}.

\bibitem{private}
F.~Staub. Private communication.

\bibitem{Cao:2013gba}
J.~Cao, F.~Ding, C.~Han, J.~M. Yang, and J.~Zhu, ``{A light Higgs scalar in the
  NMSSM confronted with the latest LHC Higgs data},''
  \href{http://dx.doi.org/10.1007/JHEP11(2013)018}{{\em JHEP} {\bfseries 1311}
  (2013) 018},
\href{http://arxiv.org/abs/1309.4939}{{\ttfamily arXiv:1309.4939 [hep-ph]}}.

\bibitem{Baron:2013eja}
{\bfseries ACME} Collaboration, J.~Baron {\em et~al.}, ``{Order of Magnitude
  Smaller Limit on the Electric Dipole Moment of the Electron},''
  \href{http://dx.doi.org/10.1126/science.1248213}{{\em Science} {\bfseries
  343} no.~6168, (2014) 269--272},
\href{http://arxiv.org/abs/1310.7534}{{\ttfamily arXiv:1310.7534
  [physics.atom-ph]}}.

\bibitem{Sakemi:2011zz}
Y.~Sakemi, K.~Harada, T.~Hayamizu, M.~Itoh, H.~Kawamura, {\em et~al.},
  ``{Search for a permanent EDM using laser cooled radioactive atom},''
\href{http://dx.doi.org/10.1088/1742-6596/302/1/012051}{{\em J.Phys.Conf.Ser.}
  {\bfseries 302} (2011) 012051}.

\bibitem{Kawall:2011zz}
D.~Kawall, ``{Searching for the electron EDM in a storage ring},''
\href{http://dx.doi.org/10.1088/1742-6596/295/1/012031}{{\em J.Phys.Conf.Ser.}
  {\bfseries 295} (2011) 012031}.

\bibitem{Kara:2012ay}
D.~Kara, I.~Smallman, J.~Hudson, B.~Sauer, M.~Tarbutt, {\em et~al.},
  ``{Measurement of the electron's electric dipole moment using YbF molecules:
  methods and data analysis},''
  \href{http://dx.doi.org/10.1088/1367-2630/14/10/103051}{{\em New J.Phys.}
  {\bfseries 14} (2012) 103051},
\href{http://arxiv.org/abs/1208.4507}{{\ttfamily arXiv:1208.4507
  [physics.atom-ph]}}.

\end{thebibliography}\endgroup

\end{document}